\documentclass[journal]{IEEEtran}
\usepackage{subfigure}
\usepackage{color}
\usepackage{cite}
\usepackage{amsmath}
\usepackage{cite}
\usepackage{graphicx}
\usepackage{epstopdf}
\usepackage{amsfonts,amsmath,amssymb}
\usepackage{graphicx}
\usepackage{url}
\usepackage{bm}
\usepackage{bbm}
\usepackage{subfigure}
\usepackage{stfloats}
\usepackage{color}

\usepackage{cite}

\newcommand{\Hnull}{\mathcal{H}_0}
\newcommand{\Halt}{\mathcal{H}_1}

\newcommand{\Honull}{\mathcal{{D}}_0}
\newcommand{\Hoalt}{\mathcal{{D}}_1}

\usepackage{cite,graphicx,amsmath,amssymb,cite,algorithm}

\newtheorem{theorem}{\textbf{Theorem}}

\newtheorem{remark}{\textbf{Remark}}

\begin{document}

\title{Optimal Detection of UAV's Transmission with Beam Sweeping in Wireless Networks}
\author{
{Jinsong Hu,~\IEEEmembership{Member,~IEEE,} Yongpeng Wu,~\IEEEmembership{Senior Member,~IEEE,} Riqing Chen, \\ Feng Shu,~\IEEEmembership{Member,~IEEE,} and  Jiangzhou Wang,~\IEEEmembership{Fellow,~IEEE}}
\thanks{J. Hu and  F. Shu are with the School of Electronic and Optical Engineering, Nanjing University of Science and Technology, Nanjing, China (Emails: \{jinsong\_hu, shufeng\}@njust.edu.cn).}

\thanks{Y. Wu is with Department of Electrical Engineering, Shanghai Jiao Tong University, Shanghai, China (Email: yongpeng.wu@sjtu.edu.cn).}

\thanks{R. Chen is with the Faculty of Computer Science and Information Engineering, Fujian Agriculture and Forestry University, Fuzhou, China (Email: riqing.chen@fafu.edu.cn).}

\thanks{J. Wang is with the School of Engineering and Digital Arts, University of Kent, Canterbury CT2 7NT, U.K. (Email: j.z.wang@kent.ac.uk).}


}
\maketitle

\vspace{-2cm}

\begin{abstract}
In this work, an detection strategy based on multiple antennas with beam sweeping is developed to detect UAV's potential transmission in wireless networks. Specifically, suspicious angle range where the UAV may present is divided into different sectors to potentially increase detection accuracy by using beamforming gain. We then develop the optimal detector and derive its detection error probability in a closed-form expression. We also utilize the Pinsker's inequality and Kullback-Leibler divergence to yield low-complex approximation for the detection error probability, based on which we obtain some significant insights on the detection performance. Our examination shows that there exists an optimal number of sectors that can minimize the detection error probability in some scenarios (e.g., when the number of measurements is limited). Intuitively, this can be explained by the fact that there exists an optimal accuracy of the telescope used to find an object in the sky within limited time period.
\end{abstract}

%
%
\vspace{-0.2cm}
\section{Introduction}
\IEEEPARstart{U}nmanned aerial vehicle (UAV) based communication as a promising technology has been extensively used in both military and civilian applications (e.g., surveillance, emergency communications) due to its advantages such as, high mobility, and low cost. UAV systems are more-effective and can be more flexibly deployed to provide on-demand coverage and
enhance capacity for emergency communications, such as, unexpected disaster, military operation \cite{Zeng2016Wireless,JointWu2018,HeChen2018UAV,Xiaobo2019UAV}. The popularity and accessibility of UAVs have seriously surged in recent years and obtaining a UAV is now possible for anybody due to its low cost, thus leads to some illegal uses of UAV. Due to its high mobility and low transmit power, it is possible for a UAV to be stealthy to enter into restricted military zones and transmit the intelligence (e.g., images) to the nearby cooperator. The detection of UAV's existence in some sensitive areas is a critical task for public and military security. However, the main challenges in the detection of UAV's with aid of traditional measures like radar are the low flight height and the small radar cross section (i.e., the size of UAV is small and the radar signal can be absorbed by the stealth UAV). Looking at the problem another way, the detection of the UAV's transmission is equal to detect of the presence of UAV in some scenarios (i.e., the UAV transmits covert message to cooperative node).

Wireless covert communication aims to enable a transmission between two users while guaranteeing a negligible detection probability of this transmission at a warden, which has been widely studied recently and examined in various scenarios.
For example, covert communication in the context of relay networks was examined in \cite{Hu2018covertrelay}, which shows that a relay can transmit confidential information to the corresponding destination covertly on top of forwarding the source's message. The authors of \cite{Biao2018Poisson} considered covert communications with a poisson field of interferers. The effect of finite blocklength on covert communications was examined in \cite{Shihao2018Delay} and the covert performance can be further enhanced by employing a artificial noise aided full-duplex receiver~\cite{Tingzhen2019WCL}.
On the other side of the coin, understanding how to prevent unauthorised wireless covert communication in order to avoid harm to our society is also of extreme importance to government and the military. To this end, the antennas array are utilized to monitor the sectors for potential UAV. We mainly tackle the optimal detection with beam sweeping for the covert transmission of UAV and what are the optimal number of sectors for beam sweeping to minimize detection error probability.


\emph{Notation:} Scalar variables are denoted by italic symbols. Bold upper and lower letters denote matrices and vectors, respectively. $\mathbb{E}[\cdot]$ denotes expectation operation. $\lfloor \cdot \rfloor $ denotes round down operation. Given a complex vector, and $||\cdot||$ denotes the norm. For a complex matrix, ${( \cdot )^H}$ refer to the conjugate transpose. $( \cdot )^{-1}$ indicates the inverse of an invertible matrix. ${{\bf{I}}_N}$ denotes the $N$-th order identity matrix.
\vspace{-0.2cm}
\section{System Model}
\subsection{Considered Scenario and Adopted Assumptions}

\begin{figure}[!t]
    \begin{center}
        \includegraphics[width=0.9\columnwidth]{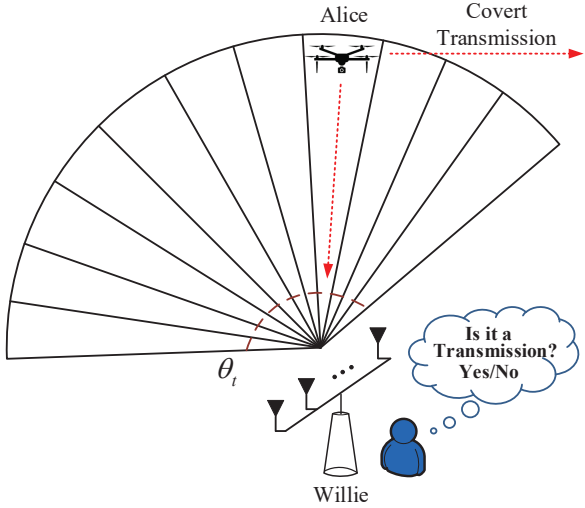}
        \caption{Detection with beam sweeping in UAV networks.}\label{fig1}
    \end{center}
    \vspace{-0.2cm}
\end{figure}

As shown in Fig.~\ref{fig1}, in this work we consider a UAV network that UAV Alice wants to spy on the target Willie and transmit covert message (e.g., intelligence information) to a cooperative UAV or a ground user under the surveillance of the warden Willie who equipped with antenna array. Due to the flight altitude, the UAV usually has line-of-sight (LoS) channel to the ground user Willie. We also assume that Alice is equipped with a single antenna. The channel from Alice to Willie is denoted by $\mathbf{h}_{aw}\in \mathbb{C}^{N\times 1}$, where $N$ is the number of antennas at Willie. In this work, we consider the 2-D scenario with polar coordinate system. The suspicious area for the surveillance UAV is denoted as a sector with the angle $\theta_t$ as shown in Fig.~\ref{fig1} and the corresponding value of phase shift for the antenna array is given by $\Theta_t=2d\cos(\theta_t)/\lambda$, where $d$ and $\lambda$ are the distance between two adjacent antennas and the carrier wavelength, respectively. Without loss of generality, we assume that the element spacing is one half wavelength, i.e., $d = \lambda/2$.

When Alice transmits covert signal in the suspicious area, the signal samples are assumed to be independent circularly symmetric zero-mean random variables with complex Gaussian distribution. We denote the hypothesis of the UAV signal being active and inactive by $\Halt$ and $\Hnull$, respectively. We assume that the additive noise samples at different antennas are independent zero-mean Gaussian random variables. Under $\Hnull$, we assume that Willie only receive the noise, while Willie will receive the UAV signal plus the noise under $\Halt$. Let $\mathbf{Y }= [\mathbf{y}_1, \cdots, \mathbf{y}_L] \in \mathbb{C}^{N\times L_t}$ be a complex matrix containing $L_t$ observed signal symbols at each antenna. The minimum angle of the sector for beam sweeping is given by \cite{Zhenyu2016Hierarchical}
\begin{align}
\Theta_{\mathrm{s}}=\frac{2}{N},
\end{align}
thus leads to the fact that $M_{\max}=\lfloor \Theta_{t}/\Theta_{\mathrm{s}}\rfloor=\lfloor N\Theta_{t}/2\rfloor$, where $M_{\max}$ is the maximum value of the number of sectors $M$. Therefore, the actual value of the angle for one sector is given by
\begin{align} \label{tilde_theta_s}
\tilde{\Theta}_{\mathrm{s}}=\frac{\Theta_{t}}{M},
\end{align}
and the corresponding number of received symbols in a sector is given by
\begin{align} \label{L_s_def}
L_s=\frac{L_t}{M} .
\end{align}

\subsection{Detection performance at Willie}
Willie has a binary detection problem, in which Alice does not transmit information in the null hypothesis $\Hnull$ but it does in the alternative hypothesis $\Halt$. The detection at Willie with multiple antennas for a sector is given by
\begin{align}
\mathbf{y}_l\sim
\left\{
  \begin{array}{ll}
    \mathcal{CN}(0,\sigma_w^2\mathbf{I}_N), &\Hnull, \\
    \mathcal{CN}(0,P_a\rho_{aw}\mathbf{h}_{aw}\mathbf{h}_{aw}^H+\sigma_w^2\mathbf{I}_N),  &\Halt,
  \end{array}
\right.
\end{align}
where $\mathbf{y}_l$ is the received symbols at Willie, $l = 1, 2, \dots, L_t$, $\sigma_w^2$ and $P_a$ are the variances of noise at Willie and UAV signal power at Alice, respectively, $\rho_{aw}\triangleq \omega (d_{aw})^{-m}$ is the pathloss between Alice and Willie, where $m$ is the path loss exponent, $\omega$ is a constant value depending on carrier frequency, which is commonly set as $[c/(4\pi f_c)]^{2}$ with $c=3~\times10^8~\mathrm{m/s}$ and $f_c$ as the carrier frequency, $d_{aw}$ is the distance from Alice to Willie.

In this work, Willie adopt the optimal detection strategy to detect the signal. The likelihood function of observation matrix $\mathbf{Y}$ under $\Hnull$ is given by
\begin{align}\label{likelihood_function_H0}
f(\mathbf{Y}| \Hnull)&=\prod_{l=1}^{L_t}\frac{1}{(\pi\sigma_w^2)^N}\exp\left\{-\frac{1}{\sigma_w^2}\mathbf{y}_l^{H}\mathbf{y}_l\right\}\\ \notag
&=\frac{1}{(\pi\sigma_w^2)^{NL_t}}\exp\left\{-\frac{1}{\sigma_w^2}\sum_{l=1}^{L_t}\mathbf{y}_l^{H}\mathbf{y}_l\right\} ,
\end{align}
where $\mathrm{tr}(\cdot)$ denotes the trace of the matrix. By taking logarithm
of \eqref{likelihood_function_H0}, i.e., $\mathcal{L}_0(\mathbf{Y})=\ln [f(\mathbf{Y}|\Hnull)]$, and using \eqref{L_s_def}, we have
\begin{align} \label{L0}
\mathcal{L}_0(\mathbf{Y})=-\frac{\mathrm{tr}(\mathbf{Y}\mathbf{Y}^H)}{\sigma_w^2}-NML_s\ln\left(\pi\sigma_w^2\right).
\end{align}

Similarly, under $\Halt$, the likelihood function can be written as
\begin{align}
&f(\mathbf{Y}|\Halt)=\prod_{l=1}^{L_s}\frac{1}{\pi^N\det(\mathbf{R})}\exp\left\{-\mathbf{y}_l^{H}\mathbf{R}^{-1}\mathbf{y}_l\right\}\times \notag \\
&~~~\prod_{k=1}^{L_t-L_s}\frac{1}{(\pi\sigma_w^2)^N}\exp\left\{-\frac{1}{\sigma_w^2}\mathbf{y}_k^{H}\mathbf{y}_k\right\} \notag \\
&=\frac{\exp\left\{-\sum_{l=1}^{L_s}\mathbf{y}_l^{H}\mathbf{R}^{-1}\mathbf{y}_l-\frac{1}{\sigma_w^2}\sum_{k=1}^{L_t-L_s}\mathbf{y}_k^{H}\mathbf{y}_k\right\}}{\pi^{NML_s}\det(\mathbf{R})^{L_s}(\sigma_w^2)^{N(M-1)L_s}},
\end{align}
where $\mathbf{R}\triangleq \mathbb{E}[\mathbf{y}_l\mathbf{y}_l^{H}|\Halt]=P_a\rho_{aw}\mathbf{h}_{aw}\mathbf{h}^H_{aw}+\sigma_w^2\mathbf{I}_{N}$ and  $\det(\mathbf{R})=(P_a\rho_{aw}||\mathbf{h}_{aw}||^2+\sigma_w^2)(\sigma_w^2)^{(N-1)}$. Then, using the matrix inversion lemma \cite{GStrang2003matrix}, we have
\begin{align}
\mathbf{R}^{-1}=\frac{1}{\sigma_w^2}\mathbf{I}-\frac{\mathbf{h}_{aw}\mathbf{h}^H_{aw}}{\left(\frac{\sigma_w^2}{P_a\rho_{aw}}+||\mathbf{h}_{aw}||^2\right)\sigma_w^2}.
\end{align}
Then, the logarithm of $f(\mathbf{Y}|\Halt)$ is given by
\begin{align} \label{L1}
&\mathcal{L}_1(\mathbf{Y})
=-\frac{\mathrm{tr}(\mathbf{Y}\mathbf{Y}^H)}{\sigma_w^2}+\frac{||\mathbf{h}^H_{aw}\mathbf{Y}||^2}{(\frac{\sigma_w^2}{P_a \rho_{aw}}+||\mathbf{h}_{aw}||^2)\sigma_w^2}- \notag\\
&~~~L_s\ln\left(\frac{P_a \rho_{aw}}{\sigma_w^2}||\mathbf{h}_{aw}||^2+1\right)-NML_s\ln(\pi\sigma_w^2).
\end{align}

Following \eqref{L0} and \eqref{L1}, the Logarithm of Likelihood Ratio (LLR) is given by
\begin{align}
\mathrm{LLR}&=\ln\left(\frac{f(\mathbf{Y};\Halt,\sigma_w^2,P_a)}{f(\mathbf{Y};\Hnull,\sigma_w^2)}\right) \notag\\
&=\mathcal{L}_1(\mathbf{Y})-\mathcal{L}_0(\mathbf{Y}) \notag\\
&=\frac{||\mathbf{h}^H_{aw}\mathbf{Y}||^2}{(\frac{\sigma_w^2}{P_a\rho_{aw}}+||\mathbf{h}_{aw}||^2)\sigma_w^2}- \notag \\
&~~~L_s\ln\left(\frac{P_a\rho_{aw}}{\sigma_w^2}||\mathbf{h}_{aw}||^2+1\right).
\end{align}
As per the LLR, the optimal decision rule is given by
\begin{align}  \label{LLR}
\frac{||\mathbf{h}^H_{aw}\mathbf{Y}||^2}{\!\left(\!\frac{\sigma_w^2}{P_a\rho_{aw}}\!+\!||\mathbf{h}_{aw}||^2\!\right)\!\sigma_w^2}\!-\!L_s\!\ln\!\left(\!\frac{P_a\rho_{aw}}{\sigma_w^2}||\mathbf{h}_{aw}||^2\!+\!1\!\right)\!\mathop{\gtrless}\limits_{\Honull}^{\Hoalt}0.
\end{align}
Following \eqref{LLR}, the optimal decision rule can be written as
\begin{align} \label{T_definition}
T\triangleq||\mathbf{h}^H_{aw}\mathbf{Y}||^2\mathop{\gtrless}\limits_{\Honull}^{\Hoalt}\eta L_s,
\end{align}
where $\Hoalt$ and $\Honull$ are the binary decisions that infer whether Alice transmits covert message or not, respectively, and $\eta$ is defined as
\begin{align}
\eta&\triangleq\ln\!\left(\!\frac{P_a\rho_{aw}}{\sigma_w^2}||\mathbf{h}_{aw}||^2+1\!\right)\!\left(\!\frac{\sigma_w^2}{P_a\rho_{aw}}+||\mathbf{h}_{aw}||^2\!\right)\!\sigma_w^2,
\end{align}
The antenna array can achieve high power gain by steering toward a given sector with narrow beam. Using \cite[Eq. (2.22)]{Constantine}, the maximum gain of antenna array is given by
\begin{align} \label{channel_gain}
||\mathbf{h}_{aw}||^2&=4\pi\frac{f(\theta,\phi)|_{\max}}{\int_0^{2\pi}\int_{\tilde{\theta}_s}^{\frac{\pi}{2}}f(\theta,\phi)\sin(\theta)\mathrm{d}\theta \mathrm{d}\phi} =\frac{2}{\tilde{\Theta}_{\mathrm{s}}},
\end{align}
where $\cos(\tilde{\theta}_s)=\tilde{\Theta}_{\mathrm{s}}$, and $f(\theta,\phi)$ represents for radiated far field of the antenna array, which is normalized to 1 in this work.

\begin{theorem}\label{theorem1}
The detection performance of Wiliie is normally measured by its detection error probability, i.e., the sum of false alarm probability $\alpha$ and miss detection probability $\beta$, which is given by
\begin{align} \label{xi}
\xi&\triangleq \alpha  + \beta \notag \\
&=1-\frac{\gamma\left[L_s,L_s\ln(1+\varphi_w)\left(1+\frac{1}{\varphi_w}\right)\right]}{\Gamma(L_s)}+ \notag \\
&~~~\frac{\gamma\left[L_s,L_s\ln(1+\varphi_w)\left(\frac{1}{\varphi_w}\right) \right]}{\Gamma(L_s)},
\end{align}
where $\varphi_w$ is the SNR at Willie, which is given by
\begin{align}
\varphi_w\triangleq \frac{P_a\rho_{aw}||\mathbf{h}_{aw}||^2}{\sigma_w^2}.
\end{align}
\end{theorem}

\begin{IEEEproof}
To evaluate the performance of detector, we compute the complementary cumulative distribution function (CCDF) of the decision statistic under $\Hnull$ and $\Halt$, respectively. Under $\Hnull$, the random vector $\mathbf{h}_{aw}^H\mathbf{Y}$ has a Gaussian distribution, i.e., $\mathbf{h}_{aw}^H\mathbf{Y}\sim \mathcal{CN}(0,||\mathbf{h}_{aw}||^2\sigma_w^2\mathbf{I}_{L_s})$. Then, from \eqref{T_definition}, the decision statistic under $\Hnull$ has the following distribution
\begin{align}
\frac{T}{||\mathbf{h}_{aw}||^2\sigma_w^2}\sim\chi_{2L_s}^2,
\end{align}
where $\chi_{2L_s}^2$ is a chi-squared random variable with $2L_s$ degrees of freedom.

Therefore, the false alarm probability $\alpha$ is easily obtained
using CCDF of $T$ as follows,
\begin{align} \label{alpha}
\alpha&=\mathcal{P}[T>\eta L_s|\Hnull]  \notag \\
&=1-\frac{\gamma\left(L_s,\frac{\eta L_s}{||\mathbf{h}_{aw}||^2\sigma_w^2}\right)}{\Gamma(L_s)},
\end{align}
where $\Gamma(L_s)=(L_s-1)!$ is the complete gamma functions and $\gamma(\cdot,\cdot)$ is the incomplete gamma function given by
\begin{align}
\gamma(n,x)=\int_0^{x}t^{n-1}e^{-t}dt.
\end{align}

Similarly, under $\Halt$, we have $\mathbf{h}_{aw}^H\mathbf{Y}\sim \mathcal{CN}(0,||\mathbf{h}_{aw}||^2(||\mathbf{h}_{aw}||^2P_a\rho_{aw}+\sigma_w^2)\mathbf{I}_{L_s})$. Then, as per \eqref{T_definition}, the decision statistic under $\Halt$
has the following distribution
\begin{align}
\frac{T}{||\mathbf{h}_{aw}||^2(||\mathbf{h}_{aw}||^2P_a\rho_{aw}+\sigma_w^2)}\sim\chi_{2L_s}^2.
\end{align}
Therefore, the miss detection probability $\beta$ is easily evaluated as
follows
\begin{align} \label{beta}
\beta&=\mathcal{P}[T\leq\eta L_s|\Halt] \notag\\
&=\frac{\gamma\left(L_s,\frac{\eta L_s}{||\mathbf{h}_{aw}||^2(||\mathbf{h}_{aw}||^2P_a\rho_{aw}+\sigma_w^2)}\right)}{\Gamma(L_s)}.
\end{align}
Utilizing the results in \eqref{alpha} and \eqref{beta}, we can achieve the expression in \eqref{xi}.
\end{IEEEproof}

The problem of minimizing the detection error probability $\xi$ in the considered system subject to a certain constraint is given by

\begin{equation}\label{P1}
\begin{aligned}
\quad \underset{M}{\min} \quad &\xi \\
\text{s. t.} \quad  & 1\leq M \leq M_{\max},
\end{aligned}
\end{equation}
Due to the complex expressions in \eqref{xi}, the optimization problem above can only be solved by numerical search.

\subsection{Special Case of Detection Performance}
In this part, we adopt this lower bound as the detection performance metric, since the expressions of $\xi$ in \eqref{xi} are too complicated to be used for further analysis. Following Pinsker's inequality, we have a lower bound on $\xi$, which is given by \cite{Shihao2018Delay}
\begin{align}\label{pinsker}
\xi\geq1-\sqrt{\frac{1}{2}\mathcal{D}(f(\mathbf{Y}|\Hnull)||f(\mathbf{Y}|\Halt))},
\end{align}
where $\mathcal{D}(f(\mathbf{Y}|\Hnull)||f(\mathbf{Y}|\Halt))$ is the Kullback-Leibler (KL) divergence from $f(\mathbf{Y}|\Hnull)$ to $f(\mathbf{Y}|\Halt)$. Then, using the chain rule of relative entropy, $\mathcal{D}(f(\mathbf{Y}|\Hnull)||f(\mathbf{Y}|\Halt))$ is given by~\cite{Shihao2018Delay}
\begin{align} \label{KL_div}
&\mathcal{D}(f(\mathbf{Y}|\Hnull)||f(\mathbf{Y}|\Halt))=L_s\left[\ln(1+\varphi_w)-\frac{\varphi_w}{1+\varphi_w}\right] \notag \\
&\overset{a}{\approx} L_s\left[\varphi_w-\frac{\varphi_w}{1+\varphi_w}\right]\overset{b}{=}\frac{4L_t(P_a\rho_{aw})^2M}{(\sigma_w^2\Theta_{t})^2+2P_a\rho_{aw}\sigma_w^2\Theta_{t}M},
\end{align}
where $\overset{a}{\approx}$ is achieved by the approximation $\ln{(1+x)}\sim x$ when $\varphi_w$ is very small, which is due to the fact that $\varphi_w$ is normally very small in order to ensure a high detection error probability at Willie\cite{Tingzhen2019WCL}. $\overset{b}{=}$ is obtained by using \eqref{tilde_theta_s}, \eqref{L_s_def}, and \eqref{channel_gain}.

\begin{remark}\label{remark1}
The value of $\mathcal{D}(f(\mathbf{Y}|\Hnull)||f(\mathbf{Y}|\Halt))$ decreases with $L_t$ for given other parameters such as $P_a$, $\sigma_w^2$.
\end{remark}

\section{Numerical Results and Discussions}
In this section, we present numerical results to examine the performance of the considered covert communications.

\begin{figure} [t!]
    \begin{center}
    \includegraphics[width=3.2in, height=2.6in]{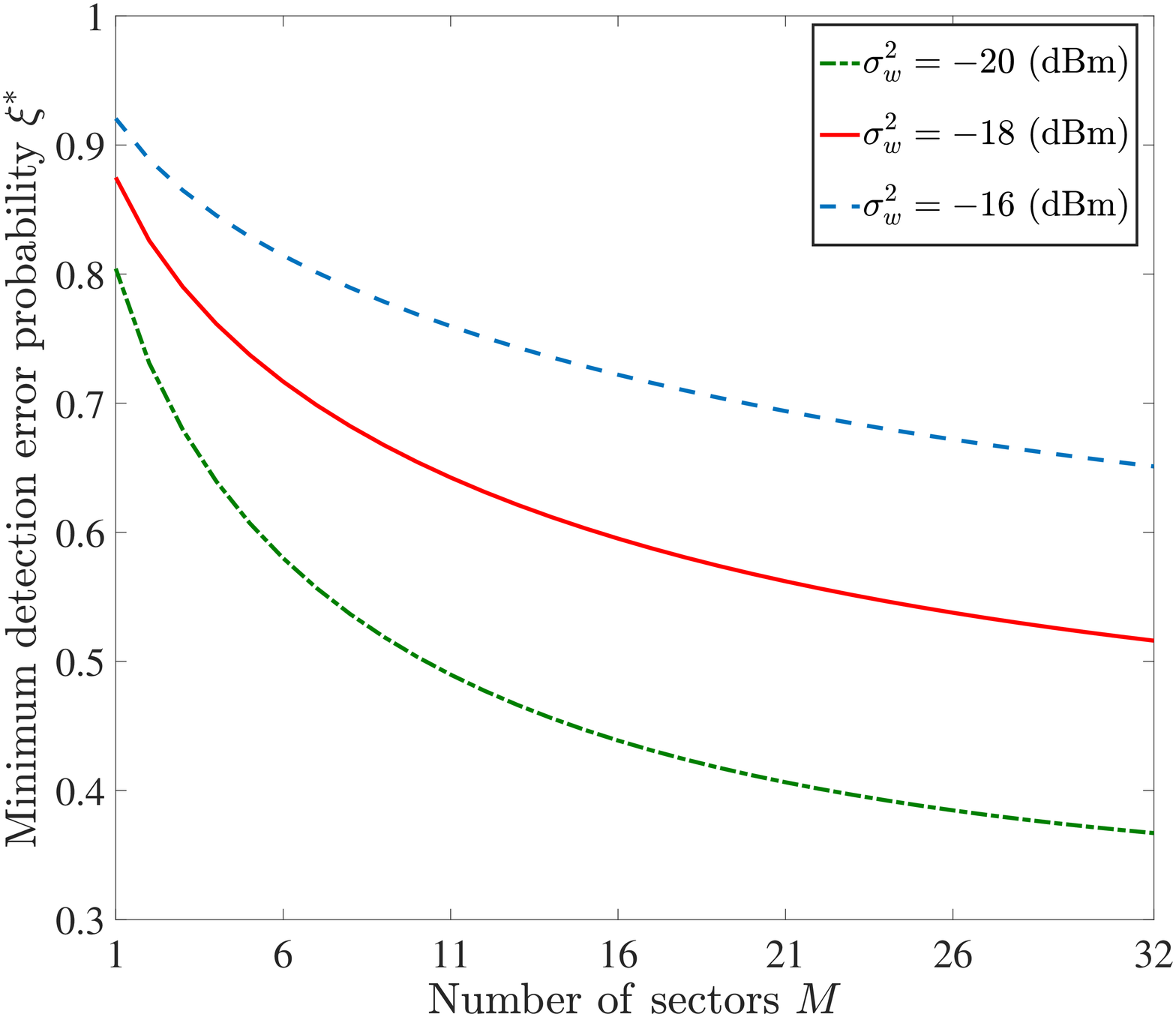}
    \caption{Minimum detection error probability $\xi$ versus the number of sectors $M$ with different values of the noise power at Willie $\sigma_w^2$, where $P_a=30$~dBm, $\Theta_t=\pi/3$, the number of antennas $N=128$, the total number of symbols $L_t=160$ and $d_{aw}=100$~m.}\label{fig2}
    \end{center}
    \vspace{-0.2cm}
\end{figure}

\begin{figure} [t!]
    \begin{center}
    \includegraphics[width=3.2in, height=2.6in]{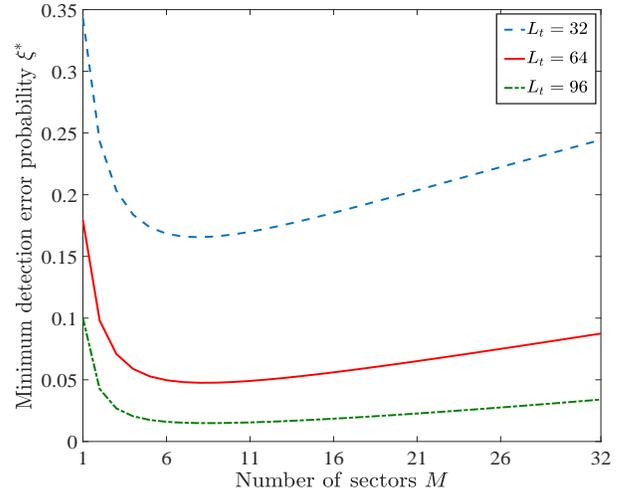}
    \caption{Minimum detection error probability $\xi$ versus the number of sectors $M$ with different values of total number of symbols $L_t$, where $P_a=10$~dBm, $\sigma_w^2=-50$~dBm, $\Theta_t=\pi/3$, the number of antennas $N=128$, and $d_{aw}=100$~m.}\label{fig3}
    \end{center}
    \vspace{-0.2cm}
\end{figure}

In Fig.~\ref{fig2}, we plot the minimum detection error probability $\xi$ versus the number of sectors $M$ with different values of the noise power at Willie $\sigma_w^2$. In this figure, we first note that $\xi$ monotonically increases as $M$ increases. We also observe that $\xi$ is a monotonically decreasing function of $\sigma_w^2$ for a given $M$, which demonstrates that the covert message becomes easier to be transmitted when $\sigma_w^2$ is larger.

Fig.~\ref{fig3} illustrates the minimum detection error probability $\xi$ versus the number of sectors $M$ with different values of the total number of symbols $L_t$. As shown in Fig.~\ref{fig3}, $\xi$ first decreases and then increases as $M$ increases. Then, we also observe that $\xi$ monotonically decreases as $L_t$ increases for a given $M$, which confirms the correctness of our Remark~\ref{remark1}. This is due to the fact that increasing $L_t$ can increase the received power at Willie.

From Fig.~\ref{fig2} and Fig.~\ref{fig3}, we find that $\xi$ monotonically increases as $M$ increases for a large $L_t$ (i.e., $L_t$=160) or $\xi$ first decreases and then increases as $M$ increases for a small value of $L_t$ (i.e., $L_t$=32). This is mainly due to the fact that the value of the received power at Willie are not limits of the detection when $L_t$ is large for a given $M$.

\vspace{-0.2cm}
\section{Conclusion}

In this work, we studied optimal detection strategy for covert communication in UAV networks with the aid of multiple antennas. Specifically, we examined the detection error possibility of using beam sweeping to detect the transmission in each sector. Our examination shows that the number of the sectors for searching the UAV's transmission has a two-side impact on the detection error possibility and setting up the number of sectors appropriately can effective reduce the detection error rate.

\vspace{-0.2cm}
\bibliographystyle{IEEEtran}
\bibliography{IEEEfull,CC}

\end{document}